# The models with phase transition for digital pixel detectors

S.V. Erin

*The state scientific center of the Russian Federation -Institute for high energy physics,*

*Research Center Kurchatov Institute, Protvino*

Abstract

The paper considers the possibility of using models with phase transition for describe the properties of digital (binary) pixel detectors without considering and taking into account the interaction between the pixels

**Introduction**

Pixel (binary) detectors - one of the major tools for obtaining "true" two-dimensional information on coordinates of particles in events with high multiplicity in high energy physics experiments. The number of pixels in modern detectors reaches several tens millions (ATLAS, CMS, ALICE). A digital detector from pixels is a set of solid-state sensors of a certain size (pixels) structurally united in a two-dimensional matrix , with the integrated read-out electronics. All the pixels in the matrix can be considered identical and they have two information states "yes " or "no".

At certain assumptions, the gas detector with pads read-out can be considered as pixel detector, especially "micro-pattern" detector. Solid-state avalanche photodiodes working in the Geiger mode (SiPMs, GAPD, MPPC, etc.) are also multi-pixel detectors [1] .

Such detectors can be considered as a set of independent cells - sensors. Of course, these pixels are not completely independent of neighboring cells due to various reasons. And this affects the parameters of the system.

The existing detectors contain approximately from $10^2$ to $10^7$ pixels. Real pixel detectors represent intermediate objects between macroscopic and microscopic systems, we will call them mesoscopic. The number of pixels in that systems is not so high that to consider the corresponding detectors as macroscopic systems, which behavior in general can be described, using average values of some quantities. Typically, the operation of a particular pixel detector is described using the moments of distribution of certain variables whose behavior depends on parameters such as the number of cells, the magnitude of the interaction between them, the threshold value.

The purpose of this work is search of the models having universal dependences which would describe properties of digital (binary) pixel detectors with the noninteracting and interacting pixels in independence of the number of pixels in the detector.

**Models of the digital (binary) pixel detector**

The digital detector consisting of identical pixels can be considered how the two-level system which is in thermal equilibrium [2]. The condition of thermal equilibrium for the pixels which are in two a state can be written down in the form of Boltzmann's distribution:

$$N_1 = N_0 \cdot \exp(-\Delta E / kT) \qquad (1)$$

where: $N_1$-the number of the noisy pixels, $N_0$ - the number of non- noisy pixels, $\Delta E$ - energy of transition from non-noisy to the noisy state, k - Boltzmann's constant, T-parameter, dimensional temperatures. Taking into account condition:

$$N = N_1 + N_0 \qquad (2)$$

(where: N is the total number of pixels), the probability of noise of the system will be:

$$P = \frac{N_1}{N} = \frac{e^{-\Delta E/kT}}{\left(1 + e^{-\Delta E/kT}\right)} \qquad (3)$$

And the difference between the number of noiseless and noisy pixels looks like this:

$$M = N_0 - N_1 = N th\left(\Delta E / 2kT\right) \qquad (4)$$

The population of each level is determined by the number of pixels the signals that exceeds or does not exceed the threshold of the readout electronics. The threshold, in turn, is the control parameter (in the case of digital signal readout) and determines the probability of a pixel p triggering, what determines the most important characteristic of any detector — its efficiency. The transition from one state to another can be interpreted as a phase transition, and the phase transition point is triggering threshold of the readout electronics. It follows from this model that when the probability of the pixel triggering is p = 1/2, the detector is in a state with maximum Shannon entropy (fig.1), that is, the pixel system is as chaotic as possible. By selecting the threshold for the operation of the readout electronics in such a way that the probability of triggering each pixel is p = 1/2, we get $N_1 = N_0$ and the detector efficiency is $\varepsilon = 50\%$. Then, by studying the behavior of the pixel distribution of the detector at this point, one can draw conclusions about the independence of the pixels from each other, or about their interaction.

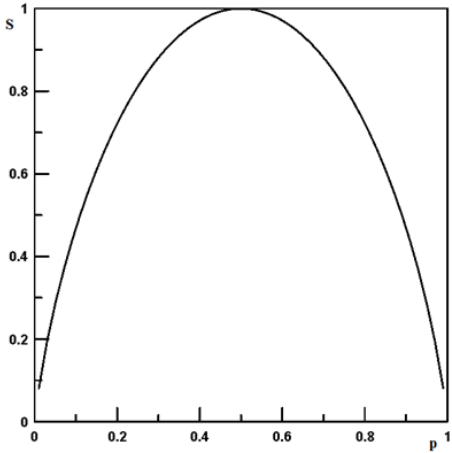

Fig.1. Dependence of entropy on the probability of finding a system in one of the states (according to Shannon)

Let's present the two-dimensional pixel detector in the form of a square lattice (fig. 2) where black rectangles represent the pixels which are in a state "1" (triggering from own noise), light rectangles represent the pixels are in a state "-1" (not initiated ). Let $p$ - the probability of finding of pixel be able "1".

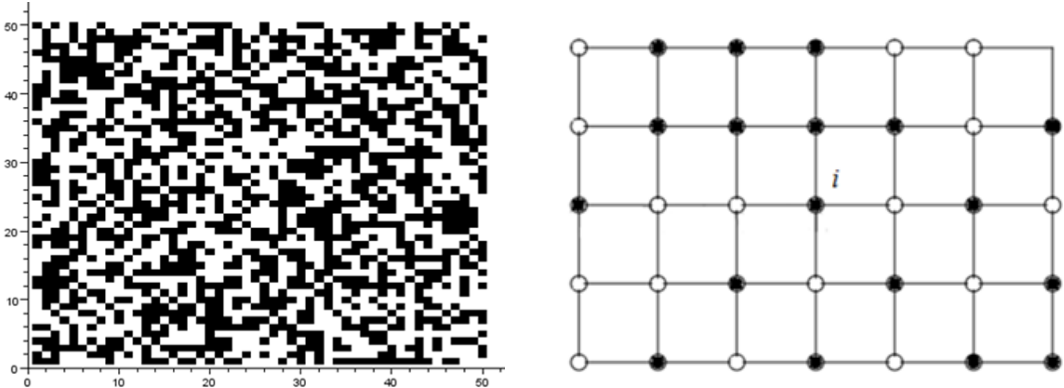

Fig. 2 Two-dimensional pixel detector with the noisy pixels and it lattice model

Consider the operation of a multi-pixel detector without external irradiation, assuming that all the triggering of the pixels is caused by its own noise. The region of activated pixels, if they are connected by a path consisting of triggered pixels, will be called a cluster [4].

Let us ask the following questions:

1.) How distribution of quantity of clusters from their sizes (quantity of busy cells) looks?

2.) Is it possible, from the information on the spatial arrangement of noisily pixels (cluster sizes, cluster size distribution), to judge the degree of interaction of pixels with each other?

3.) Is there a universal dependence that describes the properties of a multi-pixel detector, regardless of the number of pixels.

To answer these questions, let us consider models with a phase transition, which at the phase transition point have the property of scale invariance, and, in some approximation, can describe our problem. Let's list these models:

Model 1. A model with a phase transition arising in the problem of "percolation" [5] (a model of non-interacting pixels - MNP).

Model 2. A model with a phase transition arising from the use of "cellular automata" (a model of interacting pixels - MIP) [14].

Model 3. The two-dimensional Ising model [6,7].

For a model where the pixels do not interact with each other, and the triggering of each pixel occurs randomly, we will use the results of percolation theory. Similar approach was used in work [3] for studying of features of giving of working tension on detector pixels. The threshold of the electronics (the threshold for triggering a pixel) in our model corresponds to the threshold for the probability of forming a connecting cluster (the so-called percolation threshold). A connecting cluster is a cluster that connects one side of the grid to the other. The percolation threshold is model dependent on the type of lattice and the type of percolation problem [5]. In our case, it equals the probability of the pixel triggering, p, which is determined by the choice of the threshold for the operation of the electronics $Q_{por}$ (the probability of the pixel triggering and the percolation threshold are p = 1/2). Let's consider distribution of clusters by the sizes. These distributions were received as a result of a computing experiment by the Monte Carlo method (fig. 3). The matrix of 40x40 pixels in size was considered. Figure 3 shows that for p in the range from 0.45 to 0.65 (the region of the percolation threshold), the cluster size increases sharply, and with a further increase in p there are only small and maximal clusters. It is known from percolation theory that at the point of the percolation threshold the distribution of clusters is described by a power law [8,9]:

:

$$n_s \sim s^{-\tau} \tag{5}$$

where: *s*-the cluster size, an exponent $\tau \approx 2.05$ and does not depend on the number of cells in the lattice. In logarithmic scale, dependence is a straight line. According to the results given in article [9] for small cluster sizes s ~ 1 ÷ 30 and large s≥$10^5$, the behavior of the points obtained by modeling differs from the linear dependence, which is due to the finite size of the lattice. Therefore, in order to assess the degree of independence of the detector pixels, the effect of external noise, the parameters of the readout electronics on deviation from the linear dependence, it is worth choosing a grid region with a number of pixels of at least 10x10 and not more than 100x100.

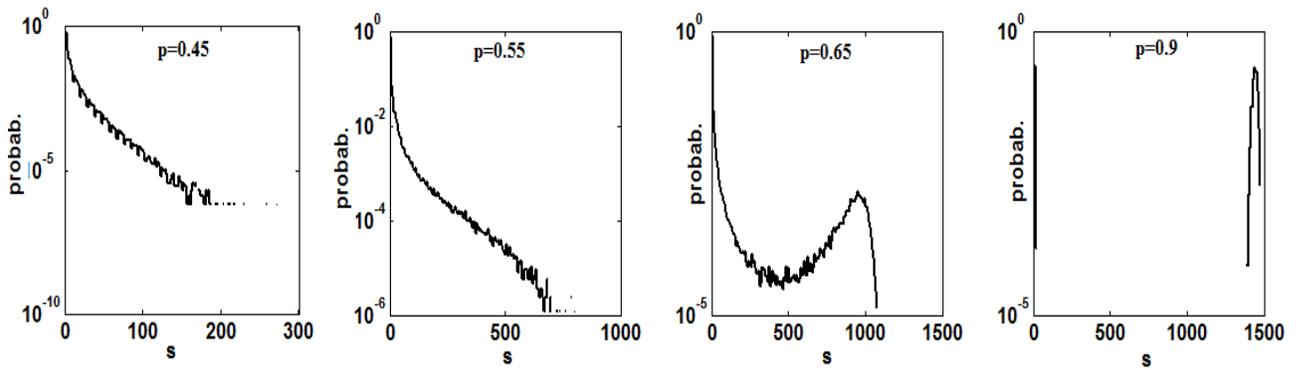

Fig.3. Probability of distribution of clusters (rated on total of clusters) from the size (quantity of cells) for various *p*.

Let's consider how the interaction between neighboring pixels affects the distribution of clusters by size. Two models can be proposed: a model based on the concept of cellular automata [10], that was first proposed von Neumann and the Ising model. First we consider the application of the model of cellular automata for our case. Suppose that each pixel interacts only with the four nearest neighbors (the so-called Neumann condition). The state of each node is determined by the states of the neighboring nodes in the previous step and its own state. The effect of the magnitude of the interaction constant between neighboring pixels - *Pcross* on the probability of cluster size distribution (without taking into account an external perturbation) was obtained as a result of a computational experiment using the model of cellular automata. The matrix of 40x40 pixels in size was considered. Cyclic boundary conditions were used. The probability of triggering a pixel was assumed to be *p* = ½. It can be seen from the figure that the shape of the distribution depends on the interaction between the pixels (Figure 4), and is a sensitive indicator of the magnitude of the interaction constant between neighboring pixels. In this model, there is a geometric phase transition, and there are two control parameters: the threshold of triggering of the read-out electronics and the probability of interaction between neighboring pixels.

To describe the characteristics of a binary pixel detector with interacting neighboring pixels in a state of thermal equilibrium (there is the possibility of a spontaneous return of the pixel to the ground state), we consider another model with a phase transition- the Ising model [11]. This model belongs to the class of lattice models, in which local interactions are considered and is widely used in many fields: in statistical physics (for investigating transitions from disorder to an ordered state), in economics or pattern recognition.

Let's again imagine a pixel detector in the form of a two-dimensional lattice (fig. 2). Consider the basis for using the Ising model. The detector has the following properties:

1.)   The initial ordered state. All pixels are not excited.

2.)   At probability of operation of pixels of p=1/2 there is a state with maximum of information entropy (fig. 1). This state can be taken as chaotic.

3.)   There is a local interaction between the next pixels.

4.)   There is a threshold of operation of pixel which defines transition of pixel from one state to other.

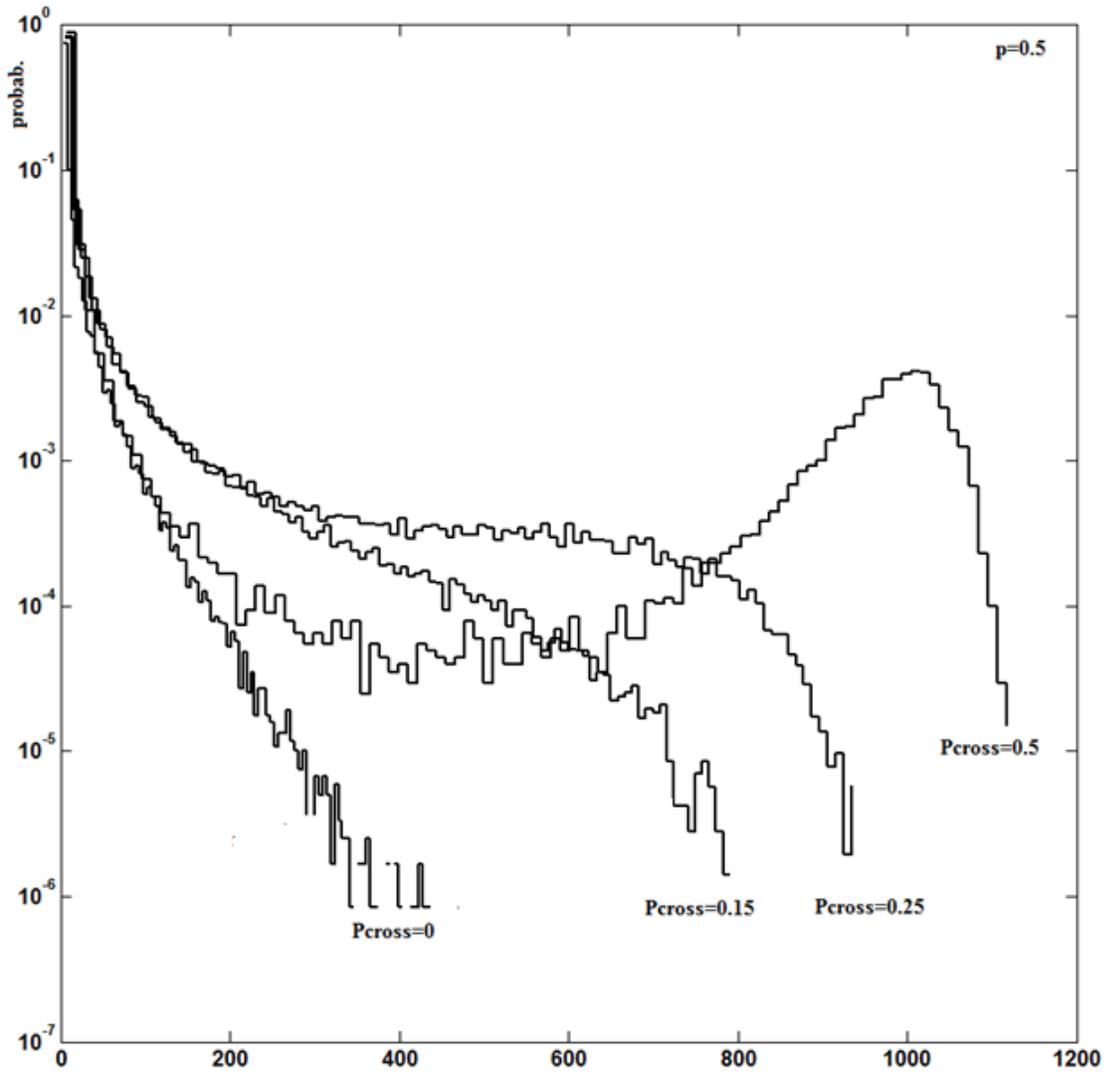

Fig.4 Probability of distribution of clusters (rated on total of clusters) from the size s (quantity of cells) for various *Pcross*.

We can say that in the process of registration, there is a transition from an ordered state to a chaotic state, which can be considered as a phase transition. According to Ising model, total energy of system looks as follows [11-14]:

$$H = H(s) = -\sum_{\langle i,j \rangle} J s_i s_j - \sum_i h s_i \qquad (6)$$

where: $s_i = \begin{cases} +1, \text{ the pixel is excited} \\ -1, \text{the pixel is not excited} \end{cases}$

the first sum undertakes on all next couples of pixels, and the second sum - on all pixels of a lattice, *h*-the energy of interaction of the external field (in our case external radiation or the general electromagnetic interference) operating on all pixels, J - is the interaction constant, which has the meaning of the binding energy between neighboring pixels. Where T is temperature, $\beta = \frac{1}{kT}$ is parameter. It is known that in the case of one-dimensional Ising model there is no phase transition, whereas in the two-dimensional one there is a phase transition at the critical temperature $T_c$ (Curie temperature) [11,12].

Magnetization of a ferromagnetic is the difference between the number of spins directed "up" and "down" [8,9]. In our case, the magnetization is the difference between the number of excited and unexcited pixels, which looks like this:

$$M = (N_0 - N_1) \qquad (7)$$

Then the detector's efficiency is related to "magnetization":

$$M = \varepsilon N - (1-\varepsilon)N$$
$$\varepsilon = \frac{1}{2} \cdot \left(1 + \frac{M}{N}\right) \qquad (8)$$

where: $\varepsilon$-average efficiency of the detector on 1 pixel, $N$ - total of pixels

It follows from expression (8) that the efficiency of the detector will be 1 if the "magnetization" is maximal, that is, all the pixels are in the unexcited state. In the case when the "magnetization" M = 0, there is the same number of excited and unexcited pixels and the efficiency $\varepsilon$ = 0.5.

Let's consider the two-dimensional pixel detector. We will discuss the model without external irradiation, since an exact analytical solution for the two-dimensional Ising model exists only for the case of the absence of an external magnetic field, and was obtained by Onsager [11]. He showed that for a two-dimensional lattice at a Curie temperature at zero external field, the ferromagnetic passes from one phase state to another:

$$T_c = \frac{2J}{\ln(1+\sqrt{2})k} \approx 2.269 \frac{J}{k} \qquad (9)$$

where is $T_c$ - Curie's temperature.

"Magnetization" at the same time is expressed by the following formula:

$$M(T) = \begin{cases}(1 - (sh(2J/kT))^{-4})^{1/8}, & T \leq T_c \\ 0, & T > T_c\end{cases} \qquad (10)$$

Thus, at Curie's temperature "magnetization" of the detector will be equal to zero, and its efficiency of equal 50%. Based on these reasons, it is possible to estimate a constant of interaction between pixels experimentally. Let the threshold of operation of the electronics, at which 50% of the detector's efficiency is achieved, will be $Q_{50\%}$ ( unit. of charge). Then it is possible to write down the following ratio connecting critical temperature with threshold value of a charge:

$$\varphi Q_{50\%} = kT_c \qquad (11)$$

where $\varphi$-the energy of formation of one electron-ion pair.

Substituting in this relation the expression for the Curie temperature, we obtain an expression for estimating the magnitude of the interaction constant between the pixels:

$$J = \frac{\varphi Q_{50\%}}{2.269} \qquad (12)$$

Let us compare the probability of cluster size distribution for the Ising model and the model of non-interacting pixels (MNP) in thermal equilibrium. The dependence of the "magnetization" on the temperature in the Ising model in the absence of an irradiation field was obtained by Monte Carlo simulation using the Metropolis algorithm [10]. Figure 5 shows that the probability of large clusters for MNPs is larger than in the Ising model, i.e. the exchange interaction reduces the size of the clusters. In our case interaction between detector pixels are lead to the excitation of neighboring pixels (which corresponds to the real model of the detector). Therefore the larger the interaction between the pixels, the smaller the critical temperature $T_c$ and the parameter value J. Thus, it is necessary to modify the Hamiltonian of the Ising model, which will allow us to more accurately find the numerical coefficient in expression (12).

The cluster size distribution in the 2D Ising model at $T = T_c$ has the following form [14]:

$$n_s \sim s^{-\tau}$$

where: is $n_s$ – is the number of clusters with size s and $\tau \approx 2.03$, regardless of the value of J and the number of pixels of the detector. Thus, the use of the dependence of the distribution of clusters on their size at fifty percent efficiency of a digital pixel detector allows us to judge the independence of pixels, the magnitude of the interaction of neighboring pixels, the thresholds of the readout electronics.

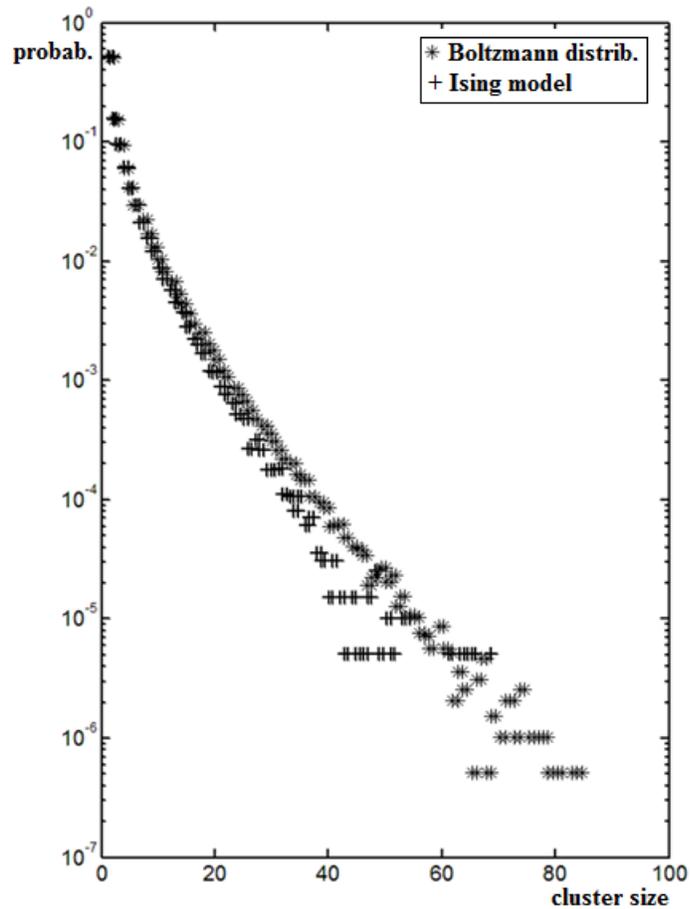

Fig.5 Probability of cluster distribution from the size (number of cells) for the Ising model and the model of noninteracting pixels at $T_c = 4 \cdot 10^7$, $Q_{50\%} = 1000e^-$, $\varphi = 3.6$ eV, $J = 0.01$ eV

**Discussion of results**

The main idea considered in this article is to find a point at which the pixels of the detector, which have a state of "1" and "0" as a result of triggering from their own noise, form a chaotic system. Then any interaction between neighboring pixels or a change in the control parameters of the system (the threshold of the recording electronics, EMI) leads to a change in the state of the system and affects the shape of the distribution of the number of clusters from their size. It turned out that to describe the behavior of such a system it is convenient to use models with a phase transition, which in this point (phase transition) are scale invariant and allow us to judge the properties of the entire system according to her part. For all the models considered, there is a universal dependence defined at the phase transition point ($\varepsilon = 50\%$), which describe the behavior of the dependence of the number of clusters on his size. This dependence has the form: $n_s \propto s^{-\tau}$, where: $n_s$ is the number of clusters with size $s$ and exponent $\tau \approx 2.03\text{-}2.05$.

In the log-log scale, this dependence is a straight line, at the ends of which, in the region of small cluster sizes and large sizes, there are small deviations caused by the finiteness of the system. The point at which the efficiency of the digital detector is $\varepsilon = 50\%$ is the point with the maximum information entropy and is the phase transition point.

Obviously, the considered models have certain limitations for describing the operation of a digital multi- pixel detector. Models 1,2 (MNP, MIP) do not contain the possibility of spontaneous return of the pixel to the ground state (recording and resetting of pixel states occurs from external signals "start", "stop"), i.e. the detector is not in thermal equilibrium with the external environment. These models allow you to easily enter the necessary parameters to describe the various pixel locations in the detector and the types of interaction between the pixels, and they should be used, for example, for silicon digital pixel detectors.

Model 3 (Ising model) may be used to describe the characteristics of a detector with interacting neighboring pixels in a state of thermal equilibrium. But as noted above, it is necessary to modify the Hamiltonian of the Ising model to refine the numerical coefficient between the interaction energy of neighboring pixels and the critical temperature. Such model can be used to describe the properties of SiPMs, GAPD, MPPC, Philips' Digital Photon Counting SiPM [15].

There are a number of practical tasks, such as: the production and testing of large-area semiconductor pixel detectors, testing of multi-pixel systems where this results can be used. The algorithm looks like this:

1.) Select any part of the detector. The sizes of cluster should lies a range $30 < s < 10^5$.

2.) Select the threshold of readout electronics. The number of triggered pixels from their own noise must be 50% of the total number of pixels.

3.) Study the distribution of the number of clusters from their size.

Based on the shape of this distribution, one can draw conclusions about the magnitude of the interaction between the pixels, the threshold of electronics, the number of operating pixels.

**Conclusion**

In this paper, it is shown that the use of information on the distribution of noise clusters by cluster size is an integral characteristic of a multi-pixel digital detector and allows one to draw conclusions about its properties.

There is a universal dependence defined at the phase transition point and describing the behavior of the dependence of the number of clusters on the size. The behavior of this dependence does not depend on the number of pixels and the magnitude of the interaction between them and has the form: $n_s \sim s^{-\tau}$, where: $n_s$ is the number of clusters with sizes s and $\tau$ in the range from 2.03 to 2.05. The phase transition point for a multi-pixel binary detector is the point at which the detector efficiency reaches 50% or the probability of triggering of the pixels $p = 1/2$, which is determined by the threshold of operation of the electronics.

## List of references